# The Phase Diagram of Carbon Dioxide from Correlation Functions and a Many-body Potential

Amanda A. Chen,[1*] Alexandria Do[1] and Tod A. Pascal[1,2*]

[1]Department of NanoEngineering and Chemical Engineering, University of California San Diego, La Jolla, CA, USA, 92023
[2]Material Science and Engineering, University of California San Diego, La Jolla, CA, USA, 92023

*corresponding authors: aac047@eng.ucsd.edu (A.A.C.), tpascal@ucsd.edu (T.A.P.)
ORCIDS: AAC: 0000-0002-7358-222X, TAP: 0000-0003-2096-1143

**ABSTRACT**

The phase stability and equilibria of carbon dioxide is investigated from 125 – 325K and 1 – 10,000 atm using extensive molecular dynamics (MD) simulations and the Two-Phase Thermodynamics (2PT) method. We devise a direct approach for calculating phase diagrams in general, by considering the separate chemical potentials of the isolated phase at specific points on the P-T diagram. The unique ability of 2PT to accurately and efficiently approximate the entropy and Gibbs energy of liquids thus allows for assignment of phase boundaries from relatively short (~ 100ps) MD simulations. We validate our approach by calculating the critical properties of the flexible Elementary Physical Model 2 (FEPM2), showing good agreement with previous results. We show, however, that the incorrect description of the short-range Pauli force and the lack of molecular charge polarization leads to deviations from experiments at high pressures. We thus develop a many-body, fluctuating charge model for $CO_2$, termed $CO_2$-Fq, from high level quantum mechanics (QM) calculations, that accurately captures the condensed phase vibrational properties of the solid (including the Fermi resonance at 1378 cm$^{-1}$) as well as the diffusional properties of the liquid, leading to overall excellent agreement with experiments over the entire phase diagram. This work provides an efficient computational approach for determining phase diagrams of arbitrary systems and underscore the critical role of QM charge reorganization physics in molecular phase stability.

## 1. INTRODUCTION

Carbon dioxide is an essential chemical, both environmentally and industrially. Human driven climate change has been largely attributed to the growing concentration of $CO_2$ in the atmosphere[1]: data from the Intergovernmental Panel on Climate Change and research studies [2, 3] indicate that $CO_2$ has the highest Radiative Forcing value, and is the greatest contributor to global warming and the greenhouse effect. Industrially, there is widespread use of supercritical carbon dioxide ($SCCO_2$), which has superior mass transfer properties, is non-toxic, cheap, and easy to recycle.[4] In heavy metal extraction, $SCCO_2$ is widely applied as the extracting solvent due to its high removal efficiency.[5] In the synthesis of Rhodium, Silver, and Copper nanoparticles, $SCCO_2$ provides a unique environment to homogenize these systems.[6-8] Additionally, $SCCO_2$ can serve as a highly selective anti-solvent in polymer synthesis, since most organic solvents show high mutual solubility with $SCCO_2$.[9]

In all these industrially and environmental processes, knowledge of the chemical and physical properties of $CO_2$ at various temperature / pressure conditions is essential, especially at extreme (high temperature and pressure) conditions. Experimental studies at these extreme conditions usually involve shock experiments,[10-12] yet these are challenging to perform in a laboratory setting as it requires highly specialized equipment. Computer simulations, employing Molecular Dynamics (MD) and/or Monte-Carlo approaches, are complementary techniques that are in principle more straightforward to perform than experiments. These simulations have been aided by the development of efficient, empirical forcefields, fitted to reproduce the properties of homogenous phases, as well as phase equilibria. Of particular note is the Elementary Physical Model 2 (EPM2),[13] which was developed to predict the liquid-vapor coexistence curve and critical properties of $CO_2$. The performance of the EPM2 over the entire phase diagram has not previously been reported, however.

Evaluating of the entire phase diagram is important since it is the ultimate metric for determining the accuracy and transferability of interaction potentials. Various computational approaches have been employed to meet this challenge, ranging from simulations in the Gibbs ensemble,[14] calculations of the latent heat across the phase boundaries and application of the Clausius-Clapeyron equation,[15] Thermodynamic Integration calculations,[16] phase-coexistence simulations,[17] and recent attempts using advanced ensemble sampling and order parameters.[18] Yet in spite of these advances, calculating the full P-T phase diagram is still a computationally expensive proposition. Moreover, while evaluation of the relative Gibbs energy of the various phases is essential, calculations of other useful thermodynamic potentials, such as entropy and heat capacity, are either difficult to obtain in the former or require additional, extensive simulations to obtain the latter.

In this study, we develop an approach for rapidly determining phase diagrams, based on explicit calculations of the entropy, enthalpy and Gibbs energy of competing phases in isolation, using the Two-phase thermodynamic (2PT) model. [19-22] The attractiveness of this approach is that it allows for the generation of the P-T phase diagram from short MD simulations, usually ~ 20ps after equilibration. Our previous work has shown that the 2PT method predicts the properties of $CO_2$ along the saturated vapor-liquid coexistence curve in good agreement with experiments, using the EPM2 model. Here, we expand on that study and show that by considering the thermodynamic properties of the $CO_2$ crystal, one can obtain good agreement compared to experiments along the entire P-T phase diagram at little extra computational cost. Moreover, our approach allows for the straightforward elaboration of the separate entropic and enthalpic energies across the phase boundaries, gaining further insights into the nature of phase transitions. Finally, we develop a new Quantum-Mechanics (QM) based, fluctuating charge forcefield, termed $CO_2$-Fq, which leads to improved performance over the phase diagram and allow us to quantify the role of intermolecular charge renormalization on phase stability.

## 2. METHODS

## 2.1 Background Theoretical Approach

2.1.1 Two-Phase Thermodynamic (2PT) Method for rapid evaluation of the Entropy and Gibbs Energy

Details of the 2PT method have been published elsewhere,[20, 21, 23] so we summarize the salient points here and direct the interested reader to our previous works [19, 24-26] and an overview in the methods sections of the supplementary materials. In 2PT, we represent the thermodynamics of a condensed phase liquid as a linear combination of two subsystems:

$Q = fQ^{gas} + (1-f)Q^{solid}$   (1)

where $Q^{gas}$ represents the thermodynamics of a hard-sphere gas, in the limit that all the modes are diffusive and $Q^{solid}$ is the thermodynamics of a Debye vibrating crystal, in the limit that all the modes are vibrational. In principle, the thermodynamic properties of these two subsystems can be obtained exactly from statistical mechanics.[27] In practice, we obtain the $Q^{gas}$ at constant density, and temperature from the Carnahan-Starling equation of state,[28, 29] while $Q^{solid}$ is obtained from the frequency dependent reweighting of the Density of States function (DoS, also known as the spectral density), as a Fourier transform of the velocity autocorrelation function [30] in MD simulations.

This superposition theory is based on early work by Eyring and Ree.[31] Lin and coworkers showed that the partition (or "fluidicity") factor *f* in Equation 1, which determines the relative weight of each subsystems, can be obtained self-consistently from an MD simulation as a ratio of the computed self-diffusion constant to that of a hard-sphere fluid at the same temperature and density.[20] When applied to molecular systems, recent work has shown that in the limit of independent molecular motions, the total system thermodynamics can be obtained from linear combinations of the thermodynamics resulting from (self-)diffusional, librational (both solid-like translations and rotations) and internal vibrational contributions.[21, 24, 26] A recent extension by Desjarlais showed that the frequency dependent distribution DoS of the gas-subsystem can be better approximated using a Gaussian memory function, which leads to improved results compared to experiments.[22]

The ultimate utility of the 2PT method in the current context is that it has the correct asymptotic behavior (by construction), so it can be applied equally in determining the thermodynamic properties of solids, liquids and gases within the same computational framework. Previous work has shown 2PT to be efficient (only required ~ 10 – 20 ps trajectories), with acceptable accuracy compared to more exact, but computationally expensive, Thermodynamic Integration and Free Energy Perturbation schemes.[23] These advantages are leveraged presently to calculate the entire phase diagram of $CO_2$, from independent MD simulations of the three competing phases at specific temperatures and pressures.

2.1.2 Partial atomic charges from the Charge Equilibration (QEq) Method

Traditional empirical forcefield are usually based on partial atomic charges that are fixed, with the total electrostatic energy obtained by Coulomb's law. While various approaches have been developed to obtain these atomic charges, in modern forcefields they are usually based on 1) population analysis of the QM wavefunction or electron density for isolated, gas-phase molecules or fragments; or 2) empirically fitted to reproduce the high order electrostatic moments (i.e., dipole, quadrupole, octupole, etc.) of the molecule. The choice of fixed atomic charges introduces some conceptual difficulties for performing simulations under conditions not explicitly considered during the charge parameterization, although modern forcefield can mitigate this somewhat by optimizing the two-body van der Waals potential. Indeed, applying this strategy, the properties of condensed phase systems using fixed charges can be reasonable under normal temperature and pressure (NTP) conditions. Specifically, in the case of $CO_2$, this strategy has led to the development of the EPM2 model, optimized to reproduce the critical properties.

Despite its attractiveness, the ability of fixed charge potentials to reliably predict the equilibrium thermodynamics far from NTP is not guaranteed and is in fact frequently compromised. This is partly due to the fact that highly compressed systems can minimize their total energy by redistributing the electron clouds around the atoms (i.e., the Pauli force), an effect that may not be correctly represented by analytic functions with power series decays, such as frequently used Lennard Jones 12-6 potentials. One approach for approximating QM charge reorganization physics is the application of polarizable forcefields. These can be generally classified as either inducible point dipole (PD),[32] classical Drude oscillators [33] or fluctuating charge (FC) [34] approaches. PD models, such as the AMOEBA [35] forcefield for example, have been successful in simulating biological systems [36], and, more recently, ionic liquids [37].

In this work, we consider FC models, due to their inherent simplicity and intuitiveness. FC models aim to address the fundamental problem of assigning partial charges to atoms within a molecule, while simultaneously minimizing the electrostatic energy, under constraints of fixed overall system charge. The most popular schemes are based on the electronegativity equalization principle of Sanderson [38], which incorporates Mulliken electronegativities [39] $\chi$ and Parr–Pearson hardness [40] $\eta$. Here, the total electrostatic energy $E(q)$ of an atom is represented as a Taylor series expansion of the charge $q$:

$$E(q) = E_0 + q\,\chi + q^2\,\eta + \dots$$

$$\chi_i = \left(\frac{\partial E}{\partial q}\right) = \frac{1}{2}(IP_i + EA_i) = -\mu_i; \quad \eta_i = \frac{1}{2}\left(\frac{\partial^2 E}{\partial q^2}\right) = \frac{1}{2}(IP_i - EA_i) \quad (2)$$

where IP is the ionization potential, EA is the electron affinity and $\mu$ is the chemical potential. The subscript $i$ represents as the $i$th atom and the difference between IP and EA represents as the idempotential, $J$. Equation 2, in effect, represents the many-body, quantum mechanical electron density in a highly simplified basis. The coulomb interactions are either calculated by means of an analytic screened coulomb function in the popular EEM scheme [41], as the overlap of Slater-type $ns$ orbitals in QEq [42-44], or more recently, as overlaps of 1s Gaussian type orbitals with atomic polarization in PQEq. [45, 46] We note that the many-body nature of FC models arises from the fact that the computed partial atomic charges are obtained self-consistently and include contributions from the self-energy as well as the interactions with other neighboring atoms. These charges are usually recalculated every step, and so varies smoothly as the local environment around the atom changes during an MD simulation. In principle, there are only two universal parameters for each element ($\chi$ and $\eta$) that can be used to reproduce the electrostatic energy of arbitrary systems. In practice, these parameters are somewhat system specific, and we present a new parameter set for $CO_2$, which we combine with various other potential energy surfaces derived from high level QM calculations, to produce the $CO_2$-Fq forcefield.

## 2.2 Computational Details

### 2.2.1. Description of Initial Systems

For simulations employing the FEPM2 forcefield, the initial structure of a $CO_2$ crystal was obtained from the ICSD [47, 48] (database code ICSD 16428), with the cubic space group 205 (Pa-3) and lattice constant a=5.624 Å. We generated a 4x4x4 supercell (256 molecules) with the initial simulation cell of 22.496 Å in x, y and z directions. To represent the liquid phase, we generated an amorphous $CO_2$ structure (256 molecules), initially at a density of 1.185 g/cm$^3$ and an initial simulation cell of 23.86 x 23.86 x 27.84 Å$^3$. For simulations of the saturated vapor/liquid at the vapor-liquid coexistence (VLE) conditions, an amorphous structure with 252 molecules was used, with initial densities obtained from the NIST database.[49] Gas systems, which are not at saturated vapor condition, contained more molecules (512 amorphous molecules) to provide enough molecule-collisions to converge the thermodynamics. When using the $CO_2$-Fq forcefield, a smaller crystal cell, with 108 $CO_2$ molecules in 3x3x3 structure (16.872 Å in x, y and z directions), was used. In the corresponding liquid simulations, we used a cell with 108 amorphous molecules at all conditions except for the VLE condition, where we used a cell with 125 molecules.

### 2.2.2. The Flexible-EPM2 Carbon Dioxide Forcefield

The FEPM2 parameters are shown in **Table 1**. The valence interactions (i.e. the C-O bond stretching and angle bending motions) are modelled as harmonic springs, which is normally sufficient to provide a similar potential energies surface compared to QM for small displacements [13]:

$$E_{valence} = E_{bonds} + E_{angles} = K_b(x - x_0)^2 + K_\Theta(\Theta - \Theta_0)^2 \quad (3)$$

where $x_0$ is the equilibrium C-O bond length and $\Theta_0$ is the equilibrium O-C-O angle. The value of the $K_b$ and $K_\Theta$ force constants are taken from our previous work [19] and Ref [13], respectively. The van der Waals interactions are described with a Lennard-Jones 12-6 potential (LJ)

$$E_{vdw}^{LJ12-6} = 4\varepsilon_{ij}\left[\left(\frac{\sigma_{ij}}{r_{ij}}\right)^{12} - \left(\frac{\sigma_{ij}}{r_{ij}}\right)^{6}\right] \quad (4)$$

with interaction energies ε and equilibrium distances σ taken from our previous work.[19]

Table 1 FEPM2 force field parameters for $CO_2$

| Atom Charge (e) | | Van der Waals (LJ) | | Bond (Harmonic) | | | Angle (Harmonic) | | |
|---|---|---|---|---|---|---|---|---|---|
| | | | ε (K) | σ (Å) | | $x_0$ (Å) | $K_b$ (kcal/mol/Å²) | $\Theta_0$ (degree) | $K_\Theta$ (kcal/mol/radian²) |
| C | 0.6512 | C | 28.13 | 2.757 | C-O | 1.149 | 1284 | O-C-O | 180 | 147.8 |
| O | -0.3256 | O | 80.51 | 3.033 | | | | | | |

2.2.3. FEPM2 Molecular Dynamics Simulations

All MD simulations were performed using the LAMMPS[50] engine. For the FEPM2 model, we initiated our simulations with 500 steps of conjugated gradient (CG) minimization. Afterwards, 10ps Langevin dynamics was applied, to heat up a system to a defined temperature. This was followed by iso-thermo / iso-baric dynamics (NPT) at the relevant pressure. To ensure equilibrium conditions, we then conducted twice 5ns of Langevin dynamics, followed by another 3ns of canonical (NVT) dynamics using a Nose-Hoover thermostat. For simulations involving the gas-phase systems, we did not perform NPT dynamics for maintaining the density property. The real space cutoffs for the Lennard-Jones and coulomb potentials were 9Å and 10Å, respectively. The long-range electrostatic were calculated with the particle-particle particle-mesh approach, with force tolerance of $10^{-4}$. We verified that this force tolerance was adequate by performing simulations with force tolerances of $10^{-6}$ and $10^{-8}$, which produced identical results.

2.2.4. FEPM2 Thermodynamics of the Solid and Liquid Phases

After equilibration, we ran an additional 200ps NVT simulation, with the trajectory (atomic positions and velocities) saved every 4fs. The thermodynamics were then obtained from an in-house code that implements the 2PT method.[51] Uncertainties in our measurements were obtained from statistical average from 10 independent simulation of 200ps each.

2.2.5. FEPM2 Thermodynamics of the Gas Phase

Two different procedures were employed to obtain the thermodynamics of the gas phases. First, we considered a low-density gas with a large number (512) of molecules and calculated the thermodynamics using the 2PT method over a 2ns sampling window. We verified that this approach has enough molecular collisions to converge the VAC and produce converged thermodynamics (Figure 1b). Secondly, we considered a high-density gas near the vapor-liquid coexistence condition, and approximated the Gibbs energies based on simulation results of the saturated vapor thermodynamics and the ideal gas equation:

$$S = S_{sat} - R\ln\left(\frac{P}{P_{sat}}\right)$$
$$E = E_{sat} \quad (5)$$

where $S$, $P$, $E$, and $R$ are denoted as entropy, pressure, internal energy, and gas constant, respectively, $S_{sat}$ and $E_{sat}$ are the entropy and internal energy respectively of a saturated vapor system at a certain temperature condition.

## 2.2.6. Construction of the CO$_2$-Fq Forcefield

We obtained the intra- / inter- molecular parameters of CO$_2$ from quantum mechanics calculations via the Q-Chem 5.2 package [52] at the aug-cc-pVTZ/MP2 level of theory. The C-O bond stretching was obtained by fitting the QM energies (**Table S1**) to a morse potential:

$$E_{bond} = D_e \left[1 - \exp(-\alpha(r - r_0))\right]^2 \quad (6)$$

with bond energy $D_e$, equilibrium distance $r_0$ and curvature $\alpha$.
The O-C-O angle bending was obtained by fitting the QM energies to a harmonic potential:

$$E_{angle} = K_\Theta (\Theta - \Theta_0)^2 \quad (7)$$

with force constant $K_\Theta$ and equilibrium angle $\Theta_0$.
The van der Waals interactions were obtained from fitting the QM binding energies of three different dimer configurations to the universal nonbonded function (UNB) [53, 54] (**Table S2**):

$$E_{vdw} = -D_e \exp\left[-\beta\left(\frac{r-R_e}{L}\right)\right] \sum_{n=0}^{5} \alpha_n \left(\frac{r-R_e}{L}\right)^n \quad (8)$$

where the $R_e$, $D_e$, and $L$ are the equilibrium distances, the binding energies, and the scaling lengths, respectively. In keeping with a previous study,[53] the parameters series ($\beta$, $\alpha_0$, $\alpha_1$, $\alpha_2$, $\alpha_3$, $\alpha_4$, $\alpha_5$) were defined as (1.00348500, 1.0, 1.02009000, 0.01678480, 0.00327294, 0.00365706, 0.00106613). We employed the UNB functional form here as it gave better fits to the QM binding energies compared to the more popular Lennard Jones, Exponential-6 or Morse potentials. The QEq parameters for carbon and oxygen in Equation 2 were fitted to reproduce the gas phase quadrupole moment of CO$_2$ from QM. The full set of parameters that defined the CO$_2$-Fq model is given in **Table 2**.

**Table 2** CO$_2$-Fq force field parameters.

|  | C - C | O - O | C – O |
|---|---|---|---|
| Van der Waals (UNB) | | | |
| $R_e$ (Å) | 5.42252 | 3.00361 | 3.64660 |
| $D_e$ (kcal/mol) | 0.04138 | 0.43302 | 0.11315 |
| $L$ (Å) | 0.69526 | 0.33056 | 0.47565 |
|  | C | O | |
| Electrostatic (QEq) | | | |
| $\chi$ (eV) | 5.34300 | 9.19962 | |
| $J$ (eV) | 10.12600 | 16.07839 | |
| $R$ (Å) | 0.75900 | 0.40344 | |
| Bond (Morse) | | | |
| $r_0$ (Å) | 1.17257 | | |
| $\alpha$ (1/Å) | 2.07474 | | |
| $D_e$ (kcal/mol) | 262.239 | | |
| Angle (Harmonic) | | | |
| $\Theta_0$ (degree) | 180 | | |
| $K_\Theta$ (kcal/mol/radian$^2$) | 55.6154 | | |

2.2.7. CO$_2$-Fq MD Solid/Liquid Simulations and Thermodynamics

In simulating the CO$_2$-Fq solid and liquid phase systems, we followed a similar procedure to section 2.2.3., expect that the system electrostatics were obtained from the overlap of Gaussian 1s orbitals via the PQEq implementation in LAMMPS (pqeq method 0). We found that application of the Generalized Langevin equation (GLE)[55, 56] to thermostat the system lead to a better distribution of energies at equilibrium. After an initial 500 steps of CG minimization, we performed 10ps of dynamics using the GLE thermostat followed by simulations in the NPT ensemble in order to stabilize the system at a specific temperature and pressure. Afterwards, we performed 3ns NVT dynamics with the GLE thermostat and another 0.5ns dynamics with the Nose-Hoover thermostat for further equilibration. The real space cutoffs for the UNB and QEq potentials were 10Å and 12.5Å, respectively, and we applied a Taper function to the QEq energies and forces to ensure zero energies and forces at the cutoff. The GLE matrix was tuned to enforce "Smart sampling",[57] with Ns=6 additional degrees of freedoms. Similar to section 2.2.4., atomic trajectory information of CO$_2$-Fq solid and liquid systems were collected for 200ps NVT dynamics. The 200ps trajectory information was further applied in 2PT thermodynamics analysis.

2.2.8. CO$_2$-Fq Thermodynamics of the Gas Phase

The QEq approach equalizes the charges in the thermodynamic limit, leading to spurious long-range charge transfer between molecules.[44, 58] This complicates simulations of gas phase systems. Thus, we obtained the CO$_2$-Fq gas-phase reference energies at specific points on the P-T diagram by applying the ideal gas law, Equation 9, and the minimized energy of the isolated molecule at 0K:

$$S = S^0 - R \ln(P/P_0)$$
$$E = E_{min} + E_{kinetic} + C_p(T - T_0) \qquad (9)$$

where the $S^0$ is the ideal gas entropy at 1atm, $E_{min}$ is the minimum energy of the system at 0 K, $E_{kinetic}$ is the kinetic energy (temperature correction), and $C_p$ is the constant pressure heat capacity correction.

2.2.9. Determination of the Phase Boundaries

We obtained the phase boundaries by explicitly considering the per-molecule Gibbs energy (i.e. the chemical potential $\mu$ for a single component system: $\mu(T,P) = g = G/N$ ) of the respective phases at specific points in the P-T diagram. The most stable phase was determined to be the one with the lowest chemical potential. We employed a multi-resolution approach to efficiently obtain the phase diagram: initial simulations were performed on a coarse sampling of the P-T space. At specific pressures, once a phase transition was detected, we first approximated the location of the phase transition temperature(s) by linear interpolation between adjacent points, followed by further simulations around this temperature in smaller temperature increments. Critical points were treated as special cases, as detailed below.

2.2.10. Determining the Critical Point from the Vapor-Liquid Coexistence curve

A variety of approaches can be used to determine the critical point, such as the discontinuity of constant pressure heat capacity, Cp,[59] or the discontinuity in relaxation times[60]. In this work, we determine the critical point via the VLE curve, [61] exploiting the fact that as the density increases, the temperature at VLE conditions will increase initially and further decrease, with the turnover point being the critical temperature.

## 3. RESULTS AND DISCUSSION

### 3.1 Spectral Density function of $CO_2$

We first tested the convergence of the 2PT method for describing the thermodynamics of $CO_2$, by considering the VAC function. Normally, for a solid, liquid, or saturated-vapor system, the VAC function converges to zero on the timescale of a few picosecond. A low-density gas, on the other hand, requires longer convergence times, due to the low collision probability between molecules. **Figure 1** presents the VAC function of $CO_2$ described by the FEPM2 and $CO_2$-Fq models, where we find convergence times of ~ 20ps for the solid and liquid systems and ~ 500ps for the gas. This result validates our computational approach, where the sampling windows (200ps and 2ns respectively) are several factors greater than these typical correlation times.

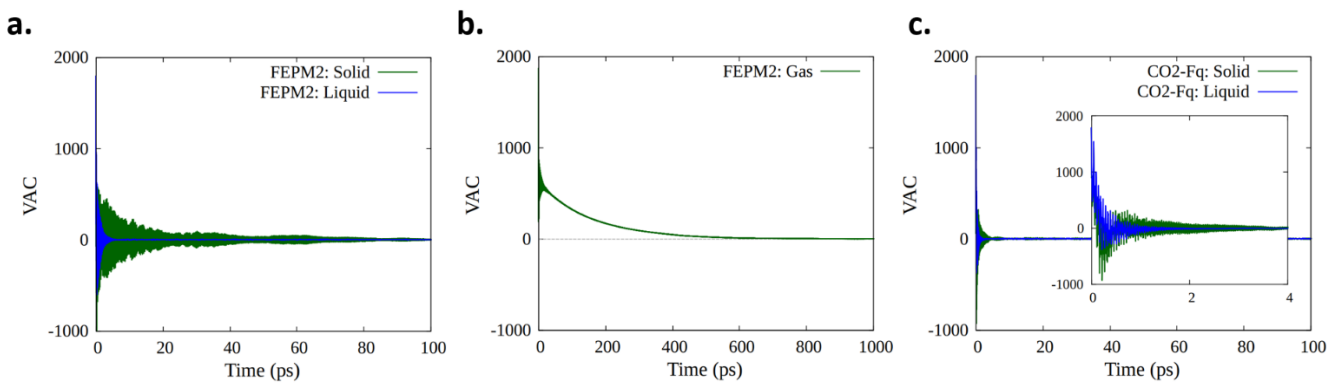

*Figure 1* The $CO_2$ total VAC function of (a) FEPM2 solid/liquid density systems at 240K,100atm, (b) FEPM2 gas density system at 1atm, 250K and (c.) $CO_2$-Fq solid/liquid density systems at 240K,100atm. The zoomed inset of (c) demonstrates the VAC function details within 0 – 4 ps.

The associated DoS of liquid $CO_2$ are shown in **Figure 2**. Here we separately consider the independent motions that contribute to the DoS: translations, rotations and internal vibrations. We apply the 2PT method to the translations and rotations, and separately show the distribution of modes from diffusive (gas-like) and from librational (solid-like) motions. The purely vibrational degrees of freedom at equilibrium are analogous to the non-equilibrium response of the system when excited by Raman and infrared radiation,[62] providing a 1:1 mapping between molecular thermodynamics and spectroscopy. We find that the vibrational spectrum of the solid and liquid phases $CO_2$ described with the FEPM2 forcefield are in reasonable agreement with the experimental asymmetric and symmetric bond stretching frequencies. However, the FEPM2 angle bending force constant (taken from the work of Trinh at el. [10]), leads to a 60% increase in the O-C-O bond bending frequency, compared to experiments (1102 cm$^{-1}$ vs 667.38 cm$^{-1}$, respectively). Additionally, the FEPM2 model is unable to reproduce the Fermi resonance peak, which results from a coupling between the O-C-O angle bending and the C-O bond stretching (i.e. cross terms), observed experimentally. Conversely, we find that the full QM-derived $CO_2$-Fq model produces improved vibrational properties compared to experiments (**Table S3**) and remarkably, captures the Fermi resonance, with two peaks at 1278 cm$^{-1}$ and 1378 cm$^{-1}$, even though the cross term was not included in the parameterization.

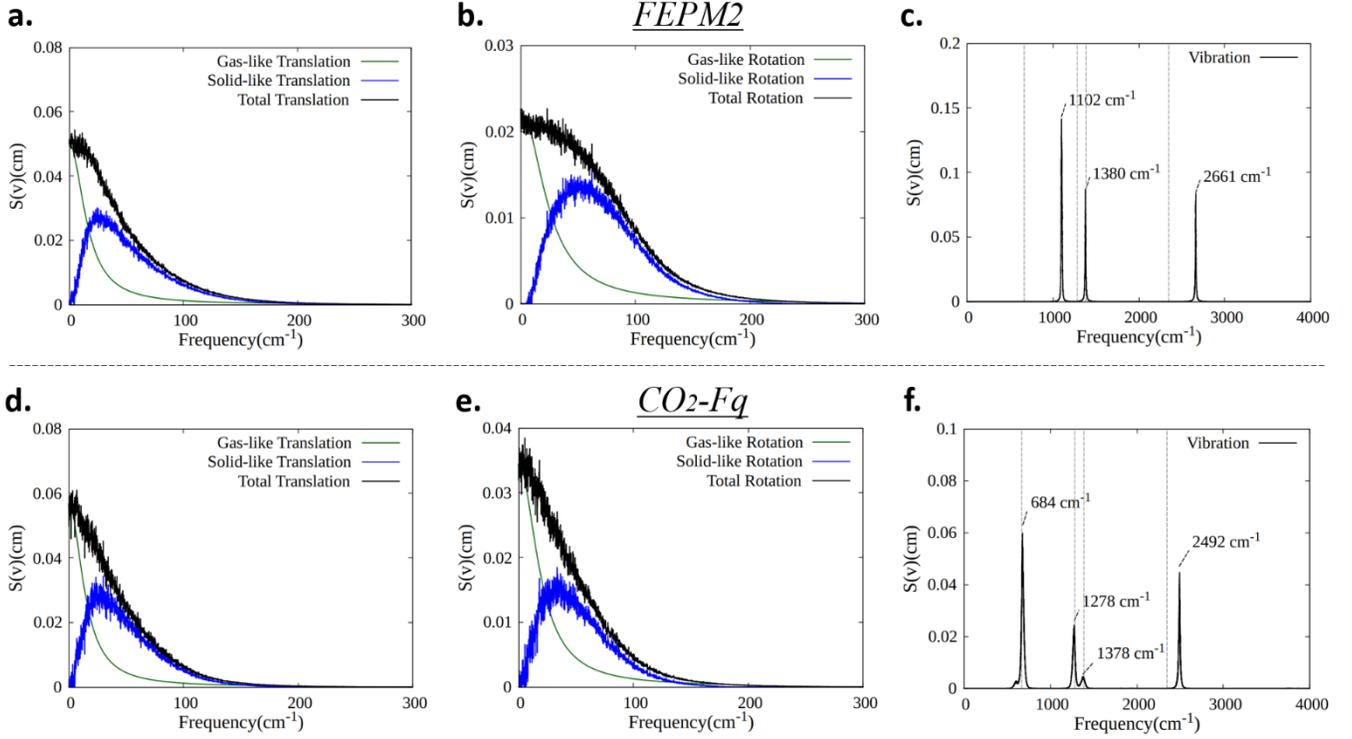

*Figure 2*. The per-molecule $CO_2$ DoS functions of (a) translational, (b) rotational, and (c) vibrational motion for FEPM2 model and the $CO_2$-Fq model (d, e and f respectively), for the liquid at 240K and 100atm. The decomposition of the translational and rotational spectrum into contributions arising from diffusive (gas-like, green) and librational (solid-like, blue) contributions as determined by the 2PT method portions are shown. The experimental vibrational frequencies are shown as dashed gray lines.

### 3.2 The Thermodynamic Properties of Carbon Dioxide

3.2.1 Thermodynamics of Crystalline $CO_2$

We now turn our attention to the thermodynamics of crystalline $CO_2$ at low temperatures, which is a more stringent test of the (gas-phase) derived interaction potentials. Specifically, we consider the calculated entropy potential, which we compare to a purely theoretical model computed from empirical parameters (**Table S4**) and Equation 10:

$$\Delta S = \int dS$$
$$= \int dH/T - \int VdP/T + \Delta H_{phase-change}/T_{phase-change}$$
$$= \int C_P^G/T\, dT + \int C_P^L/T\, dT + \int C_P^S/T\, dT - nR\int dP/P + \Delta H_{vap}/T_{vap} + \Delta H_{fus}/T_{fus} \quad (10)$$
$$\text{or}$$
$$= \int C_P^G/T\, dT + \int C_P^S/T\, dT - nR\int dP/P + \Delta h_{sub}/T_{sub}$$

where $C_p$ is the constant pressure heat capacity, the superscript *G*, *L*, and *S* represent gas, liquid and solid, respectively. *ΔH* is the phase change enthalpy, the subscript vap, fus, and sub represent vaporization, fusion, and sublimation, respectively. The number of moles of $CO_2$ in a system is denoted by *n*, *R* is the gas constant, and *P* is the pressure. We note that the 2PT entropy applies the quantum harmonic oscillator weighting function to each of the (classical) modes, a hybrid approach that produces "quantum" entropies in very good agreement with experiment for a variety of liquid systems at ambient conditions.[19, 22-26, 63-66] Of course, the 2PT method is equally applicable to purely solid and gas system, which are the limiting cases of the theory. We demonstrate this by noting that the fluidity factors for the crystalline solids are small, as expected, but slowly increasing with increasing temperature.

We find that the calculated entropies are in very good agreement with the theoretical model at low temperatures (**Table 3**). Overall, the entropy of the $CO_2$-Fq model is larger than FEPM2, reflecting the additional degree of freedom (fluctuating partial atomic charge) in the former. This ultimately leads to an overestimation of the entropy, compared to the thermodynamic model, by ~ 5 – 10%. Encouragingly, the temperature of the translational, rotational, and internal vibrational modes was consistent with the system temperatures, verifying equipartition and thermal equilibration and further validating our computational approach. Indeed, we found that application of a stochastic thermostat (Langevin or GLE) was necessary for proper mode thermal equilibration for these nanosized system on the nanosecond timescale. Application of deterministic thermostats (Nose-Hoover) required an order of magnitude longer simulation to achieve proper mode equipartition, even though the overall temperature of the system and the per-molecule distribution of total kinetic energy converged in much shorter timescale.

**Table 3:** The thermodynamic properties of crystalline $CO_2$ at 1 atm, 50K/100K/150K

| T [K] | Theoretical entropy [J/mol/K] | FEPM2 | | | $CO_2$-Fq | | |
|---|---|---|---|---|---|---|---|
| | | T Decomposition from 2PT [K] | 2PT entropy [J/mol/K] | fluidity factor | T Decomposition from 2PT [K] | 2PT entropy [J/mol/K] | fluidity factor |
| 50 | 16.19 | $T_{trans}$ = 47.07<br>$T_{rot}$ = 47.14<br>$T_{vib}$ = 53.65 | $S_q$ = 14.38 | $f_{trans}$ = 0.00327785<br>$f_{rot}$ = 0.011844 | $T_{trans}$ = 49.53<br>$T_{rot}$ = 50.54<br>$T_{vib}$ = 50.06 | $S_q$ = 20.31 | $f_{trans}$ = 0.0033372<br>$f_{rot}$ = 0.018596 |
| 100 | 37.42 | $T_{trans}$ = 96.60<br>$T_{rot}$ = 96.97<br>$T_{vib}$ = 103.85 | $S_q$ = 38.23 | $f_{trans}$ = 0.0041886<br>$f_{rot}$ = 0.021889 | $T_{trans}$ = 94.21<br>$T_{rot}$ = 96.54<br>$T_{vib}$ = 105.72 | $S_q$ = 43.94 | $f_{trans}$ = 0.0045095<br>$f_{rot}$ = 0.034896 |
| 150 | 55.92 | $T_{trans}$ = 146.21<br>$T_{rot}$ = 146.25<br>$T_{vib}$ = 154.77 | $S_q$ = 56.4 | $f_{trans}$ = 0.0053767<br>$f_{rot}$ = 0.029656 | $T_{trans}$ = 149.59<br>$T_{rot}$ = 152.43<br>$T_{vib}$ = 149.34 | $S_q$ = 63.96 | $f_{trans}$ = 0.0051085<br>$f_{rot}$ = 0.057985 |

The subscripts of trans, rot, and vib represent as the translation, rotation, and vibration modes of 2PT, respectively. $S_q$ is the quantum entropy obtained from 2PT analysis.

As a further check of equilibration, we note that the distribution of C-O bond lengths is normal and can be fit to a Gaussian function with near zero skewness (2$^{nd}$ moment) and kurtosis (3$^{rd}$ moment) (**Figure 3**).

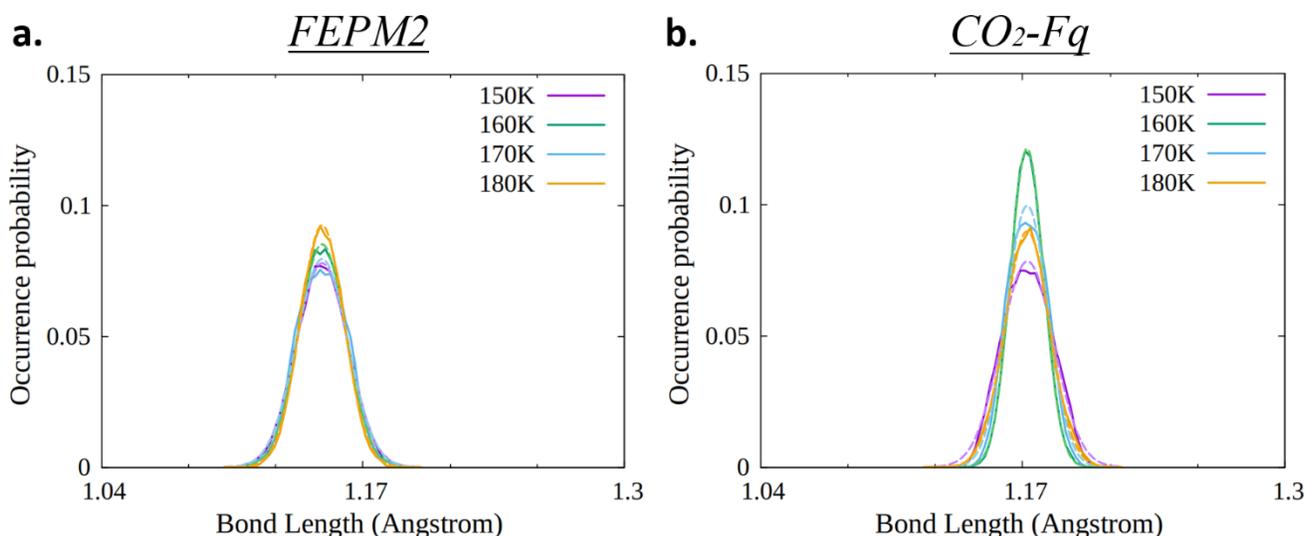

***Figure 3:*** *Probability distribution of the C-O bond lengths of crystalline $CO_2$ from equilibrium MD simulations at 150 – 180K and 1atm, using the FEPM2 (a) and $CO_2$-Fq (b) forcefields. We fit the simulation data (solid lines) to Gaussian functions (dashed lines).*

3.2.2 Carbon Dioxide Thermodynamic properties at Vapor-Liquid Coexistence (VLE) conditions and Critical Point

In **Figure 4**, we plot the density - temperature relationship of the saturated liquid and saturated vapor systems. The VLE density increases monotonically with temperature until a certain condition (i.e., critical density) is met, after which, the VLE density decreases monotonically with temperature. Thus, along the VLE curve, that saturated vapor becomes more liquid-like and saturated liquid becomes more gas-like as they approach the critical point. In fact, we note that besides the density - temperature characteristics at the VLE condition, the 2PT fluidicity-factor (f-factor) can be used to determine the critical point. This is demonstrated in **Figure 4b** for the FEPM2 model, where we find that the turning point of the curves (infinite slope) is in excellent agreement with the critical point determined from the density. At this point, the system can be described as equally liquid-like and gas-like, and the separate phases become indistinguishable.

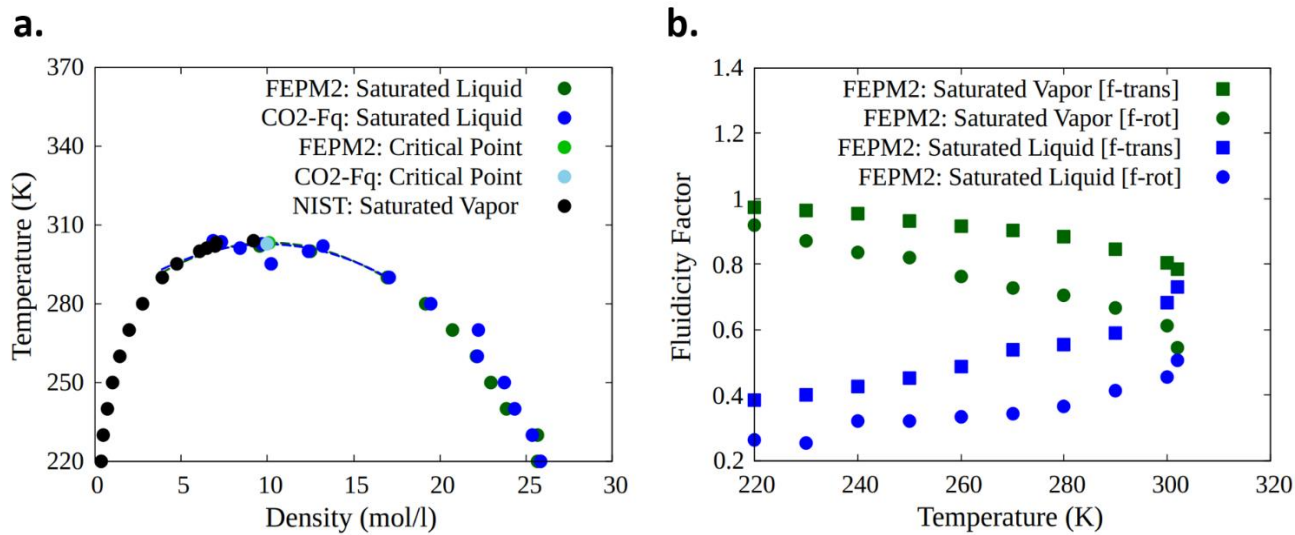

*Figure 4 a) The density – temperature relationship along the VLE curve and the critical point of $CO_2$ simulated with the FEPM2 and $CO_2$-Fq models (solid data points). Dashed curves are cubic spline fits to the calculated data. b) The translational (trans) and the rotational (rot) fluidicity factors of FEPM2 model along the VLE curve.*

Based on the results in **Figure 4**, we calculate a critical temperature ($T_c$) and a critical density ($\rho_c$) $T_c$=303.1K, $\rho_c$=10.133 mol/l for the FEPM2 model. This can be compared to Harris et al.'s work [61] ($T_c$ = 313.4 K and $\rho_c$ = 10.31 mol/l) for the original rigid EPM2 model. Further, we calculate $T_c$=302.5K, $\rho_c$=9.9883 mol/l for the $CO_2$-Fq model. Both are in a good agreement with NIST database values ($T_c$=304.18K, $d_c$=10.6 mol/l). Detailed 2PT simulated data and experimental values are shown in **Table S5**.

## 3.3 The Carbon Dioxide Phase Diagram based on phase stability and the 2PT method

As previously elaborated, we constructed the phase diagram by explicitly considering the Gibbs energy of the isolated phases at specific points on the P-T diagram. Such an approach is only possible due to the ability to compute the absolute entropy of the phases from short MD simulations using the 2PT method. This last point is important, since near the phase boundaries, we can expect significant fluctuations in the thermodynamic potentials over long-term dynamics. In fact, for 1$^{st}$ order phase transitions, the Gibbs energy function is discontinuous near the phase boundary. In **Figure 5** we plot the Gibbs energy, entropy and enthalpy of liquid/solid systems for both FEPM2 and $CO_2$-Fq models as a function of temperature at 100atm, showing the system transitions from a solid ($G^S < G^L$) to a liquid ($G^S > G^L$). We note that the FEPM2 enthalpies are computed with quantum (zero-point energy) corrections, while $CO_2$-Fq enthalpies exclude the quantum corrections for consistency with the gas phase reference.

We found that the fluctuations in the Gibbs energy were larger in the FEPM2 forcefield compared to $CO_2$-Fq, which led to larger uncertainties in the relevant phase boundaries. For example, at 100atm we were unable cleanly resolve the melting temperature ($T_m$) of the FEPM2 model by inspection, and instead had to determine $T_m$ by fitting to a cubic interpolation function, resulting in $T_m$ = 232 ± 5 K. The thermodynamics of the isolated phases are more well behaved in $CO_2$-Fq and the predicted $T_m$ = 217 ± 1 K is much better agreement with the experimental value of 218.6 K. Overall, we find a closer agreement with the experimental phase boundaries with the $CO_2$-Fq forcefield, especially in the high-pressure regime, which we attribute to an improved description of the repulsive inner wall by application of UNB nonbond potential over the Lennard Jones 12-6 potential in FEPM2.

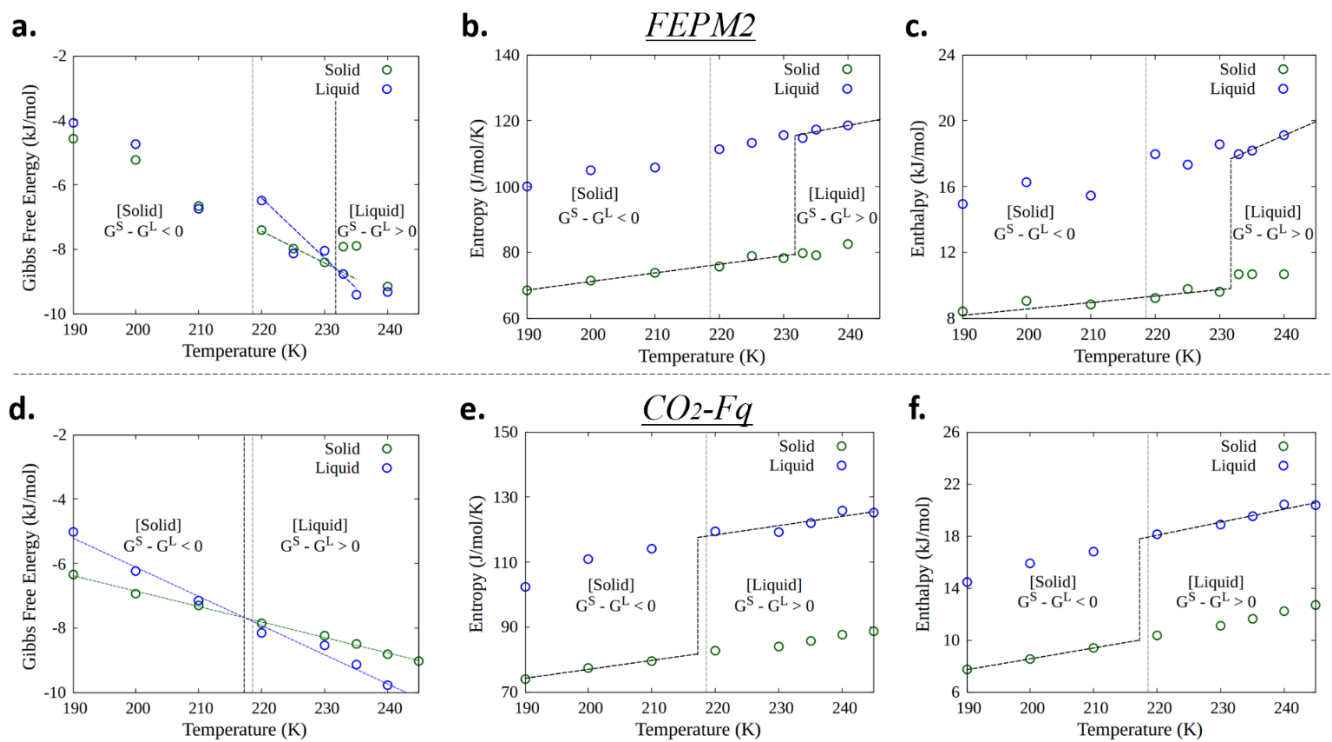

**Figure 5** *The thermodynamics of $CO_2$ described by the FEPM2 and $CO_2$-Fq models at 100atm as the system undergoes a 1st order phase transition. The FEPM2 total Gibbs energy (a), the separate entropy (b), and enthalpy (c) contributions, and the $CO_2$-Fq total Gibbs energy (d), the separate entropy (e), and enthalpy (f) contributions are shown. The dashed green line and dashed blue line represent the fitted lines in determining the phase transition temperatures via interpolation ($CO_2$-Fq) and extrapolation (FEPM2). The dashed black lines demonstrate the simulated phase changes – FEPM2 of 232 K and $CO_2$-Fq of 217 K, which can be compared to the experimental value of 218.6 K - shown as the dashed gray lines.*

In **Figure 6** we present the entire phase diagram of $CO_2$, where the phase boundaries are taken as the points in P-T space where the phases have equal chemical potentials. Here again, we note that the larger fluctuations in the Gibbs energy of FEPM2 obscures exact determination of the phase boundaries, and so we apply to interpolation scheme employed in the previous section and considered two (or more) phases to have the same chemical potentials if the difference in the Gibbs energies ($\Delta G$) are within 5%. In the case of the $CO_2$-Fq model, we were able to resolve the phase boundaries with a much more stringent condition $\Delta G < 0.3\%$.

**13**

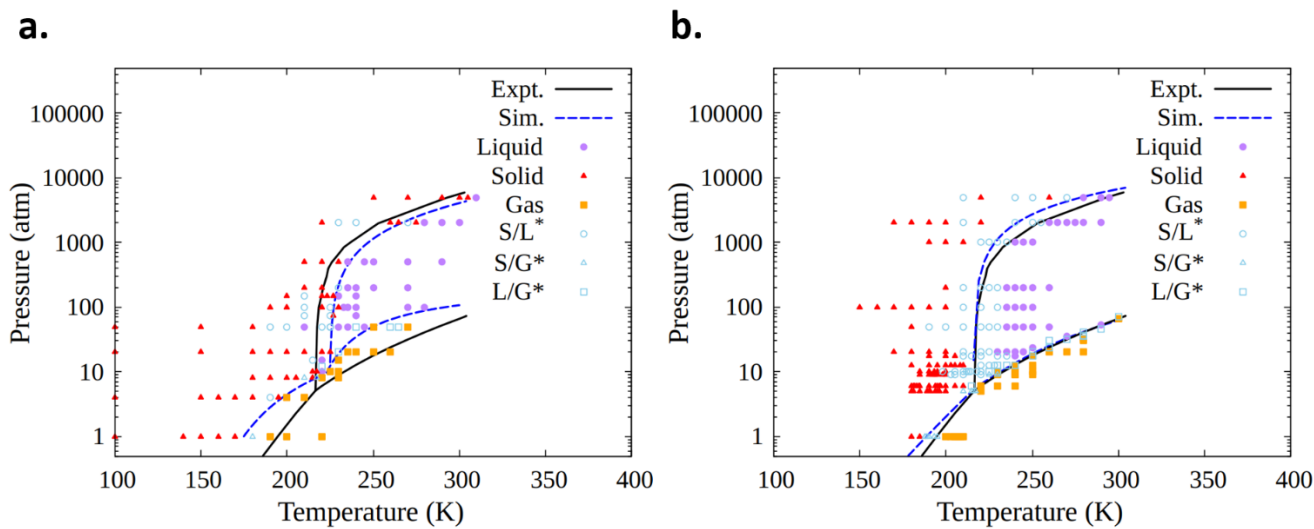

***Figure 6*** *The $CO_2$ phase diagram, based on the Gibbs energy of the isolated solid, liquid and gas phases, using the FEPM2 (a) and the $CO_2$-Fq (b) models. The superscript \* means the ΔG is smaller than 5% between solid/liquid (S/L\*), solid/gas (S/G\*) or liquid/gas (L/G\*) systems for the FEPM2, and smaller than 0.3% for the $CO_2$-Fq model. The dashed blue curves are the simulated phase boundaries of FEPM2 and $CO_2$-Fq models. The experimental reference is from the Global CCS Institute (solid black lines)*

Overall, we find improved prediction of the phase diagram using the $CO_2$-Fq model, compared to experiments. In addition to the improved van der Walls interaction mentioned previously, the additional charge degree of freedom in $CO_2$-Fq leads to more a more accurate representation of the intermolecular physics by facilitating additional instantaneous dipole interactions that stabilizes the liquid. We also obtained encouraging results when considering the triple point. Our interpolation procedure yields an approximate triple point temperature ($T_t$) and the triple point pressure ($P_t$) of $T_t = 218.0 \pm 5.0$ K, $P_t = 8.2 \pm 1.1$ atm using the FEPM2 model and $T_t = 216.2 \pm 3.0$ K, $P_t = 5.5 \pm 0.5$ atm using the $CO_2$-Fq model, both in good agreement with the experimental values from the NIST database: $T_t = 216.6$K, $P_t = 5.12$ atm. Here again, the $CO_2$-Fq model led to marked improved results compared to experiments, over FEPM2. While these results are indeed encouraging, we note that one potential limitation of the current approach is that stabilization of the isolated phases near the phase boundaries is rather difficult, even on the relatively short timescale of the 2PT trajectories. We overcome this here by interpolation near the phase boundaries, however this introduces some additional uncertainties in our calculations.

## 4. CONCLUSIONS AND OUTLOOK

In this work, we employed a fixed atom-charge model FEPM2 and a QM-derived fluctuating charge model $CO_2$-Fq to calculate the phase diagram of $CO_2$ from equilibrium MD simulations of the isolated phases and the 2PT method. We find that relatively short trajectories (~ 200ps) are sufficient for capturing the thermodynamics of the solid and liquid phases, while the gas phase require longer windows (~ 2ns). This means that the entire phase diagram was obtained from MD simulations on the ns timescale, and since the 2PT method does not incur any appreciable extra computational cost, this presents a rather efficient approach for determining phase diagram. Overall, the FEPM2 model predicts the phase behavior thermodynamics in reasonable agreement with experiments, especially at low temperatures and pressures. At higher pressures, the inability of the Lennard Jones 12-6 potential to adequately represent the repulsive Pauli forces and the inability to model charge renormalization within the molecule leads to larger deviations. The $CO_2$-Fq thus improves on the FEPM2, by including the many body QM physics, leading to excellent agreement with experiments over the entire phase diagram. This improved description may be important for study $CO_2$ in extreme environments, such as the controversial high pressure polymeric phase [67, 68] and its associated thermodynamics.

This work provides an efficient approach for calculating phase diagrams, which should be applicable to arbitrary systems. We are currently applying to approach to study other homogeneous liquids, including water. Considerations of multi-component systems are a natural extension, and insights into the behavior of the separate entropic and enthalpic functions are currently being explored.

## 5. SUPPLEMENTARY MATERIALS

See the supplementary material for complete information description of the 2PT and QEq methods and well as various tabulated energies in Tables S1 – S5.

## 6. ACKNOWLEDGEMENTS

This research was supported by NSF through the UC San Diego Materials Research Science and Engineering Center (UCSD MRSEC), DMR-2011924. T.A.P. acknowledge the start-up fund from the Jacob School of Engineering at the University of California, San Diego (UCSD). This research used resources of the National Energy Research Scientific Computing Center, a DOE Office of Science User Facility supported by the Office of Science of the U.S. Department of Energy under Contract No. DE-AC02-05CH11231. This work also used the Extreme Science and Engineering Discovery Environment (XSEDE), and the Comet and Expanse supercomputers at the San Diego Supercomputing Center, which is supported by National Science Foundation grant number ACI-1548562.

## 7. REFERENCES

[1] D. Lüthi, M. Le Floch, B. Bereiter, T. Blunier, J.-M. Barnola, U. Siegenthaler, D. Raynaud, J. Jouzel, H. Fischer, K. Kawamura, T.F. Stocker, High-resolution carbon dioxide concentration record 650,000–800,000 years before present, Nature 453 (2008) 379-382.
[2] M. Etminan, G. Myhre, E.J. Highwood, K.P. Shine, Radiative forcing of carbon dioxide, methane, and nitrous oxide: A significant revision of the methane radiative forcing, Geophysical Research Letters 43 (2016).
[3] D.S. Lee, D.W. Fahey, P.M. Forster, P.J. Newton, R.C.N. Wit, L.L. Lim, B. Owen, R. Sausen, Aviation and global climate change in the 21st century, Atmos Environ (1994) 43 (2009) 3520-3537.
[4] C. Aymonier, A. Loppinet-Serani, H. Reverón, Y. Garrabos, F. Cansell, Review of supercritical fluids in inorganic materials science, The Journal of Supercritical Fluids 38 (2006) 242-251.
[5] F. Lin, D. Liu, S. Maiti Das, N. Prempeh, Y. Hua, J. Lu, Recent Progress in Heavy Metal Extraction by Supercritical CO2Fluids, Industrial & Engineering Chemistry Research 53 (2014) 1866-1877.
[6] M. Ohde, H. Ohde, C.M. Wai, Catalytic hydrogenation of arenes with rhodium nanoparticles in a water-in-supercritical CO2 microemulsion, Chemical Communications (2002) 2388-2389.
[7] H. Ohde, F. Hunt, C.M. Wai, Synthesis of Silver and Copper Nanoparticles in a Water-in-Supercritical-Carbon Dioxide Microemulsion, Chem. Mater. 13 (2001) 4130-4135.
[8] M.J. Clarke, K.L. Harrison, K.P. Johnston, S.M. Howdle, Water in Supercritical Carbon Dioxide Microemulsions: Spectroscopic Investigation of a New Environment for Aqueous Inorganic Chemistry, J. Am. Chem. Soc. 119 (1997) 6399-6406.
[9] S.-D. Yeo, E. Kiran, Formation of polymer particles with supercritical fluids: A review, The Journal of Supercritical Fluids 34 (2005) 287-308.
[10] G.L. Schott, Shock-compressed carbon dioxide: liquid measurements and comparisons with selected models, International Journal of High Pressure Research 6 (1991) 187-200.
[11] W.J. Nellis, A.C. Mitchell, F.H. Ree, M. Ross, N.C. Holmes, R.J. Trainor, D.J. Erskine, Equation of state of shock-compressed liquids: Carbon dioxide and air, J. Chem. Phys. 95 (1991) 5268-5272.

[12] S. Root, K.R. Cochrane, J.H. Carpenter, T.R. Mattsson, Carbon dioxide shock and reshock equation of state data to 8 Mbar: Experiments and simulations, Phys. Rev. B 87 (2013) 224102.
[13] T.T. Trinh, T.J.H. Vlugt, S. Kjelstrup, Thermal conductivity of carbon dioxide from non-equilibrium molecular dynamics: A systematic study of several common force fields, J. Chem. Phys. 141 (2014) 134504.
[14] A.Z. Panagiotopoulos, Direct determination of phase coexistence properties of fluids by Monte Carlo simulation in a new ensemble, Mol Phys 61 (1987) 813-826.
[15] T. Yagasaki, M. Matsumoto, H. Tanaka, Phase Diagrams of TIP4P/2005, SPC/E, and TIP5P Water at High Pressure, J. Phys. Chem. B 122 (2018) 7718-7725.
[16] D. Frenkel, B. Smit, Understanding molecular simulation: from algorithms to applications, Academic press2001.
[17] M.M. Conde, C. Vega, Determining the three-phase coexistence line in methane hydrates using computer simulations, J. Chem. Phys. 133 (2010) 064507.
[18] P.M. Piaggi, M. Parrinello, Calculation of phase diagrams in the multithermal-multibaric ensemble, J. Chem. Phys. 150 (2019) 244119.
[19] S.-N. Huang, T.A. Pascal, W.A. Goddard III, P.K. Maiti, S.-T. Lin, Absolute Entropy and Energy of Carbon Dioxide Using the Two-Phase Thermodynamic Model, Journal of Chemical Theory and Computation 7 (2011) 1893-1901.
[20] S.T. Lin, M. Blanco, W.A. Goddard, The two-phase model for calculating thermodynamic properties of liquids from molecular dynamics: Validation for the phase diagram of Lennard-Jones fluids, J. Chem. Phys. 119 (2003) 11792-11805.
[21] S.-T. Lin, P.K. Maiti, W.A. Goddard, Two-Phase Thermodynamic Model for Efficient and Accurate Absolute Entropy of Water from Molecular Dynamics Simulations, J. Phys. Chem. B 114 (2010) 8191-8198.
[22] M.P. Desjarlais, First-principles calculation of entropy for liquid metals, Phys. Rev. E 88 (2013) 062145.
[23] C. Zhang, L. Spanu, G. Galli, Entropy of Liquid Water from Ab Initio Molecular Dynamics, J. Phys. Chem. B 115 (2011) 14190-14195.
[24] T.A. Pascal, W.A. Goddard III, Interfacial Thermodynamics of Water and Six Other Liquid Solvents, J. Phys. Chem. B 118 (2014) 5943-5956.
[25] T.A. Pascal, D. Schärf, Y. Jung, T.D. Kühne, On the absolute thermodynamics of water from computer simulations: A comparison of first-principles molecular dynamics, reactive and empirical force fields, J. Chem. Phys. 137 (2012) 244507.
[26] T.A. Pascal, S.-T. Lin, W.A. Goddard III, Thermodynamics of liquids: standard molar entropies and heat capacities of common solvents from 2PT molecular dynamics, Phys. Chem. Chem. Phys. 13 (2011) 169-181.
[27] M.S. Shell, Thermodynamics and statistical mechanics: an integrated approach, Cambridge University Press2015.
[28] G.A. Mansoori, N.F. Carnahan, K.E. Starling, J.T.W. Leland, Equilibrium Thermodynamic Properties of the Mixture of Hard Spheres, J. Chem. Phys. 54 (1971) 1523-1525.
[29] N.F. Carnahan, K.E. Starling, THERMODYNAMIC PROPERTIES OF A RIGID-SPHERE FLUID, J. Chem. Phys. 53 (1970) 600-603.
[30] P.H. Berens, D.H.J. Mackay, G.M. White, K.R. Wilson, Thermodynamics and Quantum Corrections from Molecular-Dynamics for Liquid Water, J. Chem. Phys. 79 (1983) 2375-2389.
[31] H. Eyring, T. Ree, SIGNIFICANT LIQUID STRUCTURES, VI. THE VACANCY THEORY OF LIQUIDS, Proc. Natl. Acad. Sci. U. S. A. 47 (1961) 526-537.
[32] A. Warshel, M. Kato, A.V. Pisliakov, Polarizable force fields: History, test cases, and prospects, Journal of Chemical Theory and Computation 3 (2007) 2034-2045.

[33] G. Lamoureux, B. Roux, Modeling induced polarization with classical Drude oscillators: Theory and molecular dynamics simulation algorithm, J. Chem. Phys. 119 (2003) 3025-3039.
[34] T.A. Halgren, W. Damm, Polarizable force fields, Current Opinion in Structural Biology 11 (2001) 236-242.
[35] P. Ren, J.W. Ponder, Consistent treatment of inter- and intramolecular polarization in molecular mechanics calculations, J Comput Chem 23 (2002) 1497-1506.
[36] J.W. Ponder, C. Wu, P. Ren, V.S. Pande, J.D. Chodera, M.J. Schnieders, I. Haque, D.L. Mobley, D.S. Lambrecht, R.A. DiStasio, M. Head-Gordon, G.N.I. Clark, M.E. Johnson, T. Head-Gordon, Current Status of the AMOEBA Polarizable Force Field, J. Phys. Chem. B 114 (2010) 2549-2564.
[37] O. Borodin, Polarizable force field development and molecular dynamics simulations of ionic liquids, J. Phys. Chem. B 113 (2009) 11463-11478.
[38] R.T. Sanderson, Chemical bonds and bond energy, Academic Press, New York, 1976.
[39] R.S. Mulliken, A New Electroaffinity Scale; Together with Data on Valence States and on Valence Ionization Potentials and Electron Affinities, J. Chem. Phys. 2 (1934) 782-793.
[40] R.G. Parr, R.G. Pearson, Absolute hardness: companion parameter to absolute electronegativity, J. Am. Chem. Soc. 105 (1983) 7512-7516.
[41] W.J. Mortier, K. Van Genechten, J. Gasteiger, Electronegativity equalization: application and parametrization, J. Am. Chem. Soc. 107 (1985) 829-835.
[42] A.K. Rappe, W.A. Goddard, Charge Equilibration for Molecular-Dynamics Simulations, J. Phys. Chem. 95 (1991) 3358-3363.
[43] S.W. Rick, S.J. Stuart, B.J. Berne, Dynamical fluctuating charge force fields: Application to liquid water, The Journal of Chemical Physics 101 (1994) 6141-6156.
[44] J. Chen, T.J. Martínez, QTPIE: Charge transfer with polarization current equalization. A fluctuating charge model with correct asymptotics, Chemical Physics Letters 438 (2007) 315-320.
[45] J.J. Oppenheim, S. Naserifar, W.A. Goddard III, Extension of the Polarizable Charge Equilibration Model to Higher Oxidation States with Applications to Ge, As, Se, Br, Sn, Sb, Te, I, Pb, Bi, Po, and At Elements, The Journal of Physical Chemistry A (2017).
[46] S. Naserifar, D.J. Brooks, W.A. Goddard III, V. Cvicek, Polarizable charge equilibration model for predicting accurate electrostatic interactions in molecules and solids, J. Chem. Phys. 146 (2017) 124117.
[47] G. Bergerhoff, I. Brown, F. Allen, Crystallographic databases, International Union of Crystallography, Chester 360 (1987) 77-95.
[48] D. Zagorac, H. Müller, S. Ruehl, J. Zagorac, S. Rehme, Recent developments in the Inorganic Crystal Structure Database: theoretical crystal structure data and related features, Journal of applied crystallography 52 (2019) 918-925.
[49] E. Lemmon, Thermophysical Properties of Fluid Systems, NIST chemistry WebBook, NIST standard reference database number 69, http://webbook. nist. gov. (2005).
[50] S. Plimpton, FAST PARALLEL ALGORITHMS FOR SHORT-RANGE MOLECULAR-DYNAMICS, J. Comput. Phys. 117 (1995) 1-19.
[51] https://github.com/atlas-nano/2PT
[52] Y. Shao, Z. Gan, E. Epifanovsky, A.T. Gilbert, M. Wormit, J. Kussmann, A.W. Lange, A. Behn, J. Deng, X. Feng, Advances in molecular quantum chemistry contained in the Q-Chem 4 program package, Mol Phys 113 (2015) 184-215.
[53] S. Naserifar, J.J. Oppenheim, H. Yang, T. Zhou, S. Zybin, M. Rizk, W.A. Goddard, Accurate non-bonded potentials based on periodic quantum mechanics calculations for use in molecular simulations of materials and systems, J. Chem. Phys. 151 (2019) 154111.
[54] S. Naserifar, W.A. GoddardIII, The quantum mechanics-based polarizable force field for water simulations, 149 (2018) 174502.

[55] M. Ceriotti, G. Bussi, M. Parrinello, Colored-noise thermostats à la carte, Journal of Chemical Theory and Computation 6 (2010) 1170-1180.
[56] M. Ceriotti, G. Bussi, M. Parrinello, Nuclear quantum effects in solids using a colored-noise thermostat, Phys. Rev. Lett. 103 (2009) 030603.
[57] Generated at http://cosmo-epfl.github.io/gle4md
[58] S.M. Valone, S.R. Atlas, An empirical charge transfer potential with correct dissociation limits, J. Chem. Phys. 120 (2004) 7262-7273.
[59] A.W. Nowicki, M. Ghosh, S.M. McClellan, D.T. Jacobs, Heat capacity and turbidity near the critical point of succinonitrile–water, J. Chem. Phys. 114 (2001) 4625.
[60] J.A. Lipa, C. Edwards, M.J. Buckingham, Specific heat of $CO_2$ near the critical point, Physical Review A 15 (1977) 778-789.
[61] J.G. Harris, K.H. Yung, Carbon Dioxide's Liquid-Vapor Coexistence Curve And Critical Properties as Predicted by a Simple Molecular Model, J. Phys. Chem. 99 (1995) 12021-12024.
[62] M.P. Allen, D.J. Tildesley, Computer simulation of liquids, Oxford university press2017.
[63] T.A. Pascal, C.P. Schwartz, K.V. Lawler, D. Prendergast, The purported square ice in bilayer graphene is a nanoscale, monolayer object, J. Chem. Phys. 150 (2019) 231101.
[64] B.R. Shrestha, S. Pillai, A. Santana, S.H. Donaldson Jr, T.A. Pascal, H. Mishra, Nuclear Quantum Effects in Hydrophobic Nanoconfinement, J. Phys. Chem. Lett. 10 (2019) 5530-5535.
[65] T.A. Pascal, K.H. Wujcik, D.R. Wang, N.P. Balsara, D. Prendergast, Thermodynamic origins of the solvent-dependent stability of lithium polysulfides from first principles, Phys. Chem. Chem. Phys. 19 (2017) 1441-1448.
[66] T.A. Pascal, I. Villaluenga, K.H. Wujcik, D. Devaux, X. Jiang, D.R. Wang, N. Balsara, D. Prendergast, Liquid Sulfur Impregnation of Microporous Carbon Accelerated by Nanoscale Interfacial Effects, Nano Lett. 17 (2017) 2517-2523.
[67] J. Sun, D.D. Klug, R. Martoňák, J.A. Montoya, M.-S. Lee, S. Scandolo, E. Tosatti, High-pressure polymeric phases of carbon dioxide, Proc. Natl. Acad. Sci. U. S. A. 106 (2009) 6077-6081.
[68] F. Datchi, M. Moog, F. Pietrucci, A.M. Saitta, Polymeric phase V of carbon dioxide has not been recovered at ambient pressure and has a unique structure, Proc. Natl. Acad. Sci. U. S. A. 114 (2017) E656-E657.